\documentclass[12pt]{article}

\usepackage{amsmath, amssymb}
\usepackage{amsfonts}

\begin{document}

%%%%%%%%%%%%%%%%%%%%%%%%%%%%%%%%%%%%%%%%%%%%%%%%%%%%%%%%%%%%%%%%%%%%
%%%%%%%%%%%%%%%%%%%%%%%%%%%%%%  Defs. %%%%%%%%%%%%%%%%%%%%%%%%%%%%%%
%%%%%%%%%%%%%%%%%%%%%%%%%%%%%%%%%%%%%%%%%%%%%%%%%%%%%%%%%%%%%%%%%%%%

\def\a{\alpha}
\def\b{\beta}
\def\c{\varepsilon}
\def\d{\delta}
\def\e{\epsilon}
\def\f{\phi}
\def\g{\gamma}
\def\h{\theta}
\def\k{\kappa}
\def\l{\lambda}
\def\m{\mu}
\def\n{\nu}
\def\p{\psi}
\def\q{\partial}
\def\r{\rho}
\def\s{\sigma}
\def\t{\tau}
\def\u{\upsilon}
\def\v{\varphi}
\def\w{\omega}
\def\x{\xi}
\def\y{\eta}
\def\z{\zeta}
\def\D{{\mit \Delta}}
\def\G{\Gamma}
\def\H{\Theta}
\def\L{\Lambda}
\def\F{\Phi}
\def\P{\Psi}
\def\S{\Sigma}
\def\V{\varPsi}
\newcommand{\EV}{ \,{\rm eV} }
\newcommand{\KEV}{ \,{\rm keV} }
\newcommand{\MEV}{ \,{\rm MeV} }
\newcommand{\GEV}{ \,{\rm GeV} }
\newcommand{\TEV}{ \,{\rm TeV} }

\def\o{\over}
\newcommand{\sla}[1]{#1 \llap{\, /}}
\newcommand{\beq}{\begin{eqnarray}}
\newcommand{\eeq}{\end{eqnarray}}
\newcommand{\gsim}{ \mathop{}_{\textstyle \sim}^{\textstyle >} }
\newcommand{\lsim}{ \mathop{}_{\textstyle \sim}^{\textstyle <} }
\newcommand{\vev}[1]{ \left\langle {#1} \right\rangle }
\newcommand{\bra}[1]{ \langle {#1} | }
\newcommand{\ket}[1]{ | {#1} \rangle }
\newcommand{\1}{\mbox{1}\hspace{-0.25em}\mbox{l}}

\newcommand{\cN}{{\cal N}}
\newcommand{\cB}{{\cal B}}
\newcommand{\Mpl}{M_{\mbox{\scriptsize pl}}}

%%%%%%%%%%%%%%%%%%%%%%%%%%%%%%%%%%%%%%%%%%%%%%%%%%%%%%%%%%%%%%%%%%%%

\baselineskip 0.7cm

\begin{titlepage}

\begin{flushright}
YITP-10-2 
\end{flushright}

\vskip 1.35cm
\begin{center}
{\large \bf
Nonlinearly Realized Extended Supergravity
}
\vskip 1.2cm
Izawa K.-I.$^{1,2}$, Y.~Nakai$^1$, and Ryo~Takahashi$^3$
\vskip 0.4cm

{\it $^1$Yukawa Institute for Theoretical Physics, Kyoto University,\\
     Kyoto 606-8502, Japan}

{\it $^2$Institute for the Physics and Mathematics of the Universe, University of Tokyo,\\
     Chiba 277-8568, Japan}
     
{\it $^3$Max-Planck-Institute f$\ddot{u}$r Kernphysik, Postfach 10 39
  80, 69029
Heidelberg, Germany}

\vskip 1.5cm

\abstract{
We provide nonlinear realization of supergravity with
an arbitrary number of supersymmetries by means of coset construction.
The number of gravitino degrees of freedom counts
the number of supersymmetries, which will be possibly probed
in future experiments.
We also consider goldstino embedding in the construction
to discuss the relation to nonlinear realization with rigid supersymmetries.
}
\end{center}
\end{titlepage}

\setcounter{page}{2}

\section{Introduction}

Supersymmetry (SUSY) is expected to be a basic structure of Nature
within the description in terms of effective relativistic field theory
with fundamental bosons and fermions.
It is even a leading candidate as low-energy physics just beyond
the standard model of elementary particles at the electroweak scale.

Furthermore, quantum theory with the maximal SUSY, as a hidden ingredient
broken in some way, seems unique enough as a unified theory of
elementary particles. As is well known, it is investigated under the name of
string/M theory, which has no arbitrary parameter to be defined.
In such a perspective,
SUSY may be a defining property of the basic laws in Nature.

The above point of view implies that a possible experimental discovery
of extended SUSY would constitute a rather direct evidence to elucidate
the string/M theory as a fundamental theory of Nature.
Although several phenomenological motivations are advocated
for simple SUSY, it is uncertain whether SUSY is relevant at low energy.
In fact, we have no direct experimental evidence of SUSY yet,
which means that extended SUSY is on an equal footing as simple SUSY
to be discovered in future experiments.%
\footnote{As an analogy, we may think that a possible discovery of
extra dimensions would come up with five-dimensional space-time
or yet higher-dimensional one on an equal footing. We also note that
SUSY in higher dimensions implies extended SUSY in four dimensions.}

One of the characteristic features of extended SUSY is the existence
of multiple superpartners. For definiteness, let us consider ${\cal N}=2$
SUSY in four space-time dimensions. Possible extra superpartners
to the standard model consist of "mirror" partners of gauginos,
higgsinos, quarks and leptons.%
\footnote{Various investigations have been pursued already
to consider ${\cal N}=2$ particle physics.
For a (surely incomplete) list of related literature, see Ref.\cite{N2PP}.}
In particular, multiple gravitinos seem most
compelling as an evidence of extended SUSY, since the presence of
the other mirror partners might look like mere presence of extra matter
multiplets observationally.

In this paper, we construct nonlinear realization of extended
supergravity in four dimensions. Although it is unclear how SUSY
is broken at a fundamental level, nonlinear realization possibly allows
us to investigate a manifestation of the hidden SUSY structure
behind interactions of elementary particles. The gravitino masses
turn out to determine universal interactions of the gravitinos with
matter fields. 
As an example of concrete implications, we note an exotic possibility that the
mirror gravitino may be discovered prior to the usual gravitino in the
minimal SUSY standard model.

Nonlinear realization describes low-energy effective field theory
of light degrees of freedom in a way independent of concrete
symmetry-breaking models.
The form of the effective action is determined solely
by the low-energy symmetries of the theory.
The nonlinear realization of $\mathcal{N} = 1$ global SUSY
was given by Akulov and Volkov \cite{Volkov:1973ix}
(for recent investigations, see Ref.\cite{Komargodski:2009rz}).
Their action reads
 \begin{eqnarray}
  \Gamma_{AV}= - M_s^4 \int d^4x\mbox{ det } \left(\delta^\nu_\mu - \frac{i}{M_s^4} \lambda \overleftrightarrow{\partial_\mu} \sigma^\nu\bar{\lambda} \right),
 \end{eqnarray}
where $\lambda_\alpha (x)$, $\bar{\lambda}_{\dot{\alpha}} (x)$ denote
the goldstino field and $M_s$ is the SUSY-breaking scale. The SUSY
transformation law is given by
\begin{equation}
\begin{split}
  \lambda'_{\alpha}(x') 
   &=  \lambda_{\alpha}(x)+\sqrt{2}M_s^2\xi_{\alpha}, \\
  \bar{\lambda}'_{\dot{\alpha}}(x') 
   &= \bar{\lambda}_{\dot{\alpha}}(x)+\sqrt{2}M_s^2\bar{\xi}_{\dot{\alpha}},
\end{split}\label{AV}
\end{equation}
with
\begin{equation}
\begin{split}
  x'{}^\mu 
   = x^\mu+\frac{i}{\sqrt{2}M_s^2}[\xi\sigma^\mu\bar{\lambda}(x)
                                -\lambda(x)\sigma^\mu\bar{\xi}].
\end{split}\label{AV2}
\end{equation}
Here, $\xi_{\alpha}$, $\bar{\xi}_{\dot{\alpha}}$ denote the fermionic
parameter of the global SUSY transformation.

$\mathcal{N} = 1$ local SUSY nonlinear realization was considered by
Deser and Zumino \cite{Deser:1977uq}. They constructed the theory by
reconciling the Akulov-Volkov effective action with the pure supergravity
action so that the full action has $\mathcal{N} = 1$ local SUSY
up to the second order of the fields. The super-Higgs mechanism results in the gravitino mass,
\begin{eqnarray}
m_{\frac{3}{2}} = \frac{M_s^2}{\sqrt{3}M_{\text{pl}}},
\label{gravitinomass}
\end{eqnarray}
where $M_{\text{pl}}$ is the reduced Planck mass scale.
We henceforth adopt the Planck unit $M_{\text{pl}}=1$.
In a similar manner, the case of extended SUSY was considered by
Ferrara, Maiani, and West \cite{Ferrara:1983fi}. They discussed the
extended local SUSY nonlinear realization up to the second order just
like the $\mathcal{N} = 1$ case by Deser and Zumino.

In contrast, coset construction of nonlinear realization developed by
Callan, Coleman, Wess, and Zumino \cite{Callan:1969sn}
does not involve such an approximation
and is valid to all orders with
regard to the fields contained in the action.
Therefore, we employ this technique to construct nonlinearly realized
extended supergravity, which contains gravitinos in addition to
ordinary matters without the need for their superpartners.
The case with $\mathcal{N} = 1$ local SUSY is considered by Clark, Love,
Nitta, and ter Veldhuis \cite{Clark:2005qu}. Our construction is a
generalization of theirs to the case with extended SUSY.

The rest of the paper is organized as follows.
In the next section, we introduce nonlinearly realized extended
supergravity by means of coset construction technique.
In section 3, we discuss goldstino embedding in our locally
SUSY theories.
The final section concludes the paper with some comments on future directions.

\section{Coset Construction}

In this section, we perform the coset construction of nonlinearly
realized supergravity with an arbitrary number of SUSY.
We first recapitulate coset construction for a general case of global symmetries.

Here some definitions follow.
$G$ is the symmetry group of internal and space-time symmetries of a theory.
For example, in our case, $G$ is given by the extended SUSY algebra,
\begin{equation}
\begin{split}
  &[M_{\mu\nu},M_{\rho\sigma}]=-i(\eta_{\mu\rho}M_{\nu\sigma}
                              -\eta_{\mu\sigma}M_{\nu\rho}
                              +\eta_{\nu\sigma}M_{\mu\rho}
                              -\eta_{\nu\rho}M_{\mu\sigma}), \\
  &[M_{\mu\nu},P_\lambda]=i(P_\mu\eta_{\nu\lambda}-P_\nu\eta_{\mu\lambda}), \\
  &[M_{\mu\nu},Q_{A\alpha}]=-\frac{1}{2}(\sigma^{\mu\nu})_\alpha^{\,\,\,\,\beta} Q_{A\beta},
 \\
  &[M_{\mu\nu},\bar{Q}_{A\dot{\alpha}}]
    =\frac{1}{2}(\bar{\sigma}^{\mu\nu})_{\dot{\alpha}}^{\,\,\,\,\dot{\beta}}
     \bar{Q}_{A\dot{\beta}}, \\
  &\{Q_{A\alpha},\bar{Q}_{B\dot{\alpha}}\}
    =2\sigma^\mu_{\alpha\dot{\alpha}}\delta_{AB}P_\mu, \\
  &\{Q_{A\alpha},Q_{B\beta}\}=\epsilon_{\alpha\beta}X_{AB}, \\
  &\{\bar{Q}_{A\dot{\alpha}},\bar{Q}_{B\dot{\beta}}\}
    =\epsilon_{\dot{\alpha}\dot{\beta}}\bar{X}_{AB},
\end{split}\label{algebra}
 \end{equation}
with the other commutators vanishing and 
$\eta_{\mu\nu}=\mbox{diag}[+1,-1,-1,-1]$.

The generators of the group $G$ can be divided into the following three sets:
\begin{itemize}
\item $P_\mu$ : the generators of space-time translations,
\item $\Gamma_a$ : the generators of spontaneously broken internal and space-time symmetries,
\item $\Gamma_i$ : the generators of unbroken internal symmetries
and space-time rotations, which form a subgroup $H$ of $G$.
\end{itemize}
Then, we define the coset space $G/H$ with 
the equivalence relation $\Omega \sim \Omega h$
for $\Omega \in G$ and $h \in H$.
A representative element of the coset space $G/H$ is written as
\begin{eqnarray}
\Omega = \exp[ix_\mu P^\mu]\exp[iN^a(x)\Gamma_a],
\end{eqnarray}
where $x_\mu$ denote space-time coordinates and $N^a(x)$ are generalized
Nambu-Goldstone fields
which contain goldstinos associated with spontaneously broken global SUSY.

The action of a group element $g$ of $G$ on a coset space element
as $\Omega \to \Omega'$ is defined by
\begin{eqnarray}
g\Omega = \Omega' h,
\end{eqnarray}
where
\begin{eqnarray}
\Omega' = \exp[ix'_\mu P^\mu]\exp[i{N'}^a(x') \Gamma_a],
\end{eqnarray}
and $h = \exp[i\alpha^i(g,x,N) \Gamma_i]\in H$ with $\alpha^i(g,x,N)$
as a function of $g$, $x$, and $N$. That is, the induced
transformations of $x_\mu$ and $N^a(x)$ as
\begin{eqnarray}
x_\mu \to x'_\mu, \quad N^a(x) \to {N'}^a(x')
\end{eqnarray}
are given through the group action.

In order to construct invariant actions under the symmetry group $G$,
we now proceed to introduce the Maurer-Cartan 1-form,
which is a Lie algebra valued 1-form defined by
\begin{eqnarray}
\Omega^{-1} d\Omega = i(\omega^\mu P_\mu + \omega^a \Gamma_a + \omega^i \Gamma_i),
\end{eqnarray}
where $d$ denotes the space-time exterior derivative.
The transformation of the Maurer-Cartan 1-form under $G$ turns out to be
\begin{eqnarray}
\Omega^{-1} d\Omega \to h(\Omega^{-1} d\Omega)h^{-1} + hdh^{-1}.
\end{eqnarray}
Notice that only the $h$ appears in the above expression.
This property is crucial to
construct invariant actions of
nonlinearly realized extended supergravity below.

\subsection{The transformation laws}

Now we further discuss basic building blocks and their
transformation properties to construct invariant actions of nonlinearly
realized extended supergravity, taking into account the local nature
of SUSY transformations in supergravity.

We adopt the extended SUSY algebra given in the above Eq.\eqref{algebra}
with $H$ as the Lorentz group. 
Then, an element of the coset space $G/H$ is written as
\begin{eqnarray}
  \Omega(x)=e^{ix_\mu P^\mu}
            e^{i[\lambda_A^\alpha(x)Q_{A\alpha}
                +\bar{\lambda}_{A\dot{\alpha}}(x)\bar{Q}_A^{\dot{\alpha}}
                +C_{AB}(x)X_{AB}+\bar{C}_{AB}(x)\bar{X}_{AB}]},
  \label{Omega}
\end{eqnarray}
where the indices $A$ and $B$ run from $1$ to $\mathcal{N}$, that is,
the number of SUSY in four space-time dimensions,
and $\lambda_A^\alpha(x)$ are the goldstino fields associated with
$\mathcal{N}$ SUSY. The $C_{AB}(x)$ and $\bar{C}_{AB}(x)$,
which are antisymmetric in $A$ and $B$, are the Nambu-Goldstone fields
associated with $\mathcal{N}(\mathcal{N}-1)$ central charges.

The action of an element $g$ of the group $G$ to $G/H$ is given 
in the same way as for the global case by
\begin{eqnarray}
g\Omega = \Omega' h,
\end{eqnarray}
though, in the local case with $x$-dependent $g$, we have
\begin{equation}
\begin{split}
 g(x)&=e^{i\epsilon^\mu(x)P_\mu}
       e^{i[\xi_A^\alpha(x)Q_{A\alpha}+\bar{\xi}_{A\dot{\alpha}}(x)\bar{Q}_A^{\dot{\alpha}}]}
       e^{i[\zeta_{AB}(x)X_{AB}+\bar{\zeta}_{AB}(x)\bar{X}_{AB}]}
       e^{\frac{i}{2}\alpha^{\mu\nu}(x)M_{\mu\nu}},\\
 \Omega'(x') 
  &= e^{ix'{}^\mu P_\mu}
      e^{i[\lambda_A'{}^\alpha(x')Q_{A\alpha}
           +\bar{\lambda}_{A\dot{\alpha}}'(x')\bar{Q}_A^{\dot{\alpha}}
           +C_{AB}'(x')X_{AB}+\bar{C}_{AB}'(x')\bar{X}_{AB}]}, \\
  h(x) &= e^{\frac{i}{2}\alpha^{\mu\nu}(x)M_{\mu\nu}},
\end{split}
\end{equation}
where $\epsilon^\mu(x)$, $\xi_A^\mu(x)$, $\zeta_{AB}(x)$, and
$\alpha^{\mu\nu}(x)$ denote local infinitesimal parameters.
 
The Baker-Campbell-Hausdorff formula, 
 \begin{eqnarray}
  e^Xe^Y=e^{X+Y+\frac{1}{2}[X,Y]
                        +\frac{1}{12}([X,[X,Y]+[Y,[Y,X]])+\cdots},
 \end{eqnarray}
yields the nonlinear extended local SUSY transformation laws of
the generalized Nambu-Goldstone fields, which are given by
 \begin{equation}
 \begin{split}
  \lambda'_{A\alpha}(x') 
   &= \lambda_{A\alpha}(x)+\xi_{A\alpha}(x)
          +\frac{i}{4}\alpha_{\mu\nu}(x)(\sigma^{\mu\nu})_\alpha^{\,\,\,\,\beta}
           \lambda_{A\beta}(x), \\
  \bar{\lambda}'_{A\dot{\alpha}}(x') 
   &= \bar{\lambda}_{A\dot{\alpha}}(x)+\bar{\xi}_{A\dot{\alpha}}(x)
          -\frac{i}{4}\alpha_{\mu\nu}(x)(\bar{\sigma}^{\mu\nu})_{\dot{\alpha}}^{\,\,\,\,\dot{\beta}}
           \lambda_{A\dot{\beta}}(x), \\
  C_{AB}'(x')
   &= C_{AB}(x)+\zeta_{AB}(x)-\frac{i}{2}\xi_{[A}\lambda_{B]}, \\
  \bar{C}_{AB}'(x')
   &= \bar{C}_{AB}(x)+\bar{\zeta}_{AB}(x)
         -\frac{i}{2}\bar{\xi}_{[A}\bar{\lambda}_{B]},
\end{split}\label{NGtr}
\end{equation}
with
 \begin{equation}
 \begin{split}
  x'{}^\mu 
   = x^\mu+\Delta x^\mu
   \equiv x^\mu+\epsilon^\mu(x)+i[\xi_A(x)\sigma^\mu\bar{\lambda}_A(x)
                                -\lambda_A(x)\sigma^\mu\bar{\xi}_A(x)]-\alpha^{\mu\nu}(x)x_\nu,
\end{split}\label{NGtr2}
\end{equation}
where the square bracket $[AB]$ for the indices
indicates their antisymmetrization.
These constitute extended local SUSY generalization of the nonlinear
transformation law Eq.\eqref{AV} proposed by Akulov and Volkov.

How about the Maurer-Cartan form?
We would like to maintain the transformation property of the Maurer-Cartan
1-form in the global case, though
we now consider the local transformation $g(x)$, 
which results in a deviated transformation property
of $\Omega^{-1} d\Omega$.
Therefore, we define our locally covariant Maurer-Cartan 1-form through
replacing the exterior derivative by a covariant derivative, that is,
\begin{equation}
\begin{split}
  \omega &=      \Omega^{-1}D\Omega \equiv \Omega^{-1}(d+iE)\Omega \\
         &= i\left[\omega^mP_m+\frac{1}{2}\omega_{QA}^\alpha Q_{A\alpha}
                    +\frac{1}{2}\bar{\omega}_{\bar{Q}A\dot{\alpha}}\bar{Q}_A^{\dot{\alpha}}
                    +\omega_{XAB}X_{AB}+\bar{\omega}_{\bar{X}AB}\bar{X}_{AB}
                    +\frac{1}{2}\omega_M^{mn}M_{mn}\right],
\end{split}
\end{equation}
where the indices $m,n=0,1,2,3$ are hereafter used for the tangent space local Lorentz
transformation, and 
\begin{eqnarray}
  E=\hat{E}^mP_m+\frac{1}{2}\psi_A^\alpha Q_{A\alpha}+\frac{1}{2}\bar{\psi}_{A\dot{\alpha}}\bar{Q}_A^{\dot{\alpha}}
    +A_{AB}X_{AB}+\bar{A}_{AB}\bar{X}_{AB}+\frac{1}{2}\gamma^{mn}M_{mn}
 \end{eqnarray}
is the 1-form gravitational field that is also a Lie algebra valued 1-form.
As we see later, $\hat{E}^m$ represents the graviton,
$\psi_A^\alpha$, $\bar{\psi}_{A\dot{\alpha}}$ are the gravitinos,
$A_{AB}$, $\bar{A}_{AB}$ are the gauge fields associated with the
central charges, and $\gamma^{mn}$ is the spin connection.
Here, the 1-form gravitational field transforms as a gauge field:
\begin{eqnarray}
  E'(x')=g(x)E(x)g^{-1}(x)-ig(x)dg^{-1}(x).
 \end{eqnarray}
Then, the form of the transformation law for
this modified Maurer-Cartan 1-form
is the same as that for the original one, namely,
\begin{eqnarray}
  \omega'(x')=h(x)\omega(x)h^{-1}(x)+h(x)dh^{-1}(x).
 \end{eqnarray}

We can extract the transformation laws of the components of Maurer-Cartan 1-form
by expanding the above transformation in terms of the charges of the symmetry group $G$.
The transformation laws of the component 1-form fields are given by
\begin{equation}
\begin{split}
\omega'{}^m(x')     
   &= \omega^n(x)\Lambda_n^m(\alpha(x)), \\
  \omega'_{QA\alpha}(x') 
   &= D_{\alpha}^{(\frac{1}{2},0)\beta}(\alpha(x))\omega_{QA\beta}, \\
  \bar{\omega}_{\bar{Q}A}'{}^{\dot{\alpha}}(x') 
   &= D^{(0,\frac{1}{2})\dot{\alpha}}_{\qquad\,\dot{\beta}}(\alpha(x))
       \bar{\omega}_{\bar{Q}A}^{\dot{\beta}}, \\
  \omega_{XAB}'(x') &= \omega_{XAB}(x), \\
  \bar{\omega}_{\bar{X}AB}'(x') &= \bar{\omega}_{\bar{X}AB}(x), \\
  \omega_M'{}^{mn}(x')
   &= \omega_M^{rs}(x)\Lambda_r^m(\alpha(x))\Lambda_s^n(\alpha(x))
       -d\alpha^{mn}(x),
\end{split}\label{Maurertr}
\end{equation}
where the infinitesimal local Lorentz transformation and the
corresponding spinor transformation matrices are denoted by
\begin{equation}
 \begin{split}
  \Lambda_n^m(\alpha(x)) &=\delta_n^m-\alpha_n^m(x), \\
D_\alpha^{(\frac{1}{2},0)\beta}(\alpha(x)) 
   &= \delta_\alpha^\beta+\frac{i}{4}\alpha_{mn}(x)(\sigma^{mn})_\alpha^{\,\,\,\,\beta}, \\
  D^{(0,\frac{1}{2})\dot{\alpha}}_{\qquad\,\dot{\beta}}(\alpha(x))
   &= \delta^{\dot{\alpha}}_{\dot{\beta}}+\frac{i}{4}\alpha_{mn}(x)
       (\bar{\sigma}^{mn})^{\dot{\alpha}}_{\,\,\,\,\dot{\beta}}.
 \end{split}
\end{equation}
The above expressions show that the components of
locally covariant Maurer-Cartan 1-form are subject
to the appropriate local Lorentz transformations for their indices.

The gauge transformation of the gravitational 1-form
can be also expanded by the charges of the symmetry group $G$.
The transformation laws of the components are given by
\begin{equation}
\begin{split}
  \hat{E}'{}^m 
   &= \hat{E}^m+\gamma^{mn}\epsilon_n
       +i(\xi_A\sigma^m\bar{\psi}_A-\psi_A\sigma^m\bar{\xi}_A)
       -\alpha^{mn}\hat{E}_n-d\epsilon^m, \\
  \psi_A'{}^\alpha 
   &=\psi_A^\alpha
       -\frac{i}{4}\alpha_{mn}(\psi_A\sigma^{mn})^\alpha
       +\frac{i}{2}\gamma_{mn}(\xi_A\sigma^{mn})^\alpha-2d\xi_A^\alpha, \\
  \bar{\psi}_{A\dot{\alpha}}'
   &= \bar{\psi}_{A\dot{\alpha}}
       -\frac{i}{4}\alpha_{mn}(\bar{\psi}_A\bar{\sigma}^{mn})_{\dot{\alpha}}
       +\frac{i}{2}\gamma_{mn}(\bar{\xi}_A\bar{\sigma}^{mn})_{\dot{\alpha}}
       -2d\bar{\xi}_{A\dot{\alpha}}, \\
  A_{AB}' &= A_{AB}-\frac{i}{2}\xi_{[A}\psi_{B]}-d\zeta_{AB}, \\
  \bar{A}_{AB}'  
   &= \bar{A}_{AB}-\frac{i}{2}\bar{\xi}_{[A}\bar{\psi}_{B]}-d\bar{\zeta}_{AB}, \\
  \gamma'{}^{mn} 
   &= \gamma^{mn}+(\alpha^{mr}\gamma_r^n-\alpha^{nr}\gamma_r^m)
       -d\alpha^{mn}.
 \end{split}
\end{equation}
which amount to the usual transformation laws of the vierbein,
gravitino, and spin connection in supergravity.

We finally expand the locally covariant Maurer-Cartan 1-form $\omega$
by the charges of $G$ to obtain
 \begin{equation}
 \begin{split}
  \omega^m 
   &= dx^m-i[\lambda_A\sigma^m(d\bar{\lambda}_A+\bar{\psi}_A)
              -(d\lambda_A+\psi_A)\sigma^m\bar{\lambda}_A]+E^m  \\
   &\quad+\frac{1}{4}\gamma_{rs}\lambda_A(\sigma^m\bar{\sigma}^{rs}
                                       +\sigma^{rs}\bar{\sigma}^m)
        \bar{\lambda}_A, \\
  \omega_{QA}^\alpha 
   &= 2d\lambda_A^\alpha+\psi_A^\alpha
       -\frac{i}{2}\gamma_{mn}(\lambda_A\sigma^{mn})^\alpha, \\
  \bar{\omega}_{\bar{Q}A\dot{\alpha}} 
   &= 2d\bar{\lambda}_{A\dot{\alpha}}+\bar{\psi}_{A\dot{\alpha}}
       -\frac{i}{2}\gamma_{mn}(\bar{\lambda}_A\bar{\sigma}^{mn})_{\dot{\alpha}}, \\
  \omega_{XAB} 
   &= \frac{i}{2}\lambda_{[A}d\lambda_{B]}+dC_{AB}+\frac{i}{2}\lambda_{[A}\psi_{B]}+A_{AB}
       +\frac{1}{8}\gamma_{mn}\lambda_{[A}\sigma^{mn}\lambda_{B]}, \\
  \bar{\omega}_{\bar{X}AB}
   &= \frac{i}{2}\bar{\lambda}_{[A}d\bar{\lambda}_{B]}+d\bar{C}_{AB}+\frac{i}{2}\bar{\lambda}_{[A}\bar{\psi}_{B]}
       +\bar{A}_{AB}
       +\frac{1}{8}\gamma_{mn}\bar{\lambda}_{[A}\bar{\sigma}^{mn}\bar{\lambda}_{B]}, \\
  \omega_M^{mn} &= \gamma^{mn},
\end{split}
 \end{equation}
where $E^m = \hat{E}^m - \gamma^{mn} x_n$.
In the next subsection, we construct invariant actions under $G$ by
using the locally covariant Maurer-Cartan 1-form and its transformation law obtained above.

\subsection{The invariant actions}

We now proceed to construct gauge-invariant actions under the
symmetry group $G$ and express them in terms of the component fields.
Let us first describe the components $\omega^m$ of locally
covariant Maurer-Cartan 1-form by the space-time coordinate differential as
$\omega^m=dx^\mu e_\mu^m$, where $e_\mu^m$ is the "vierbein" of extended
local SUSY given by
\begin{equation}
\begin{split}
  e_\mu^m &= \delta_\mu^m-2i\lambda_A\overleftrightarrow{\partial_\mu}\sigma^m\bar{\lambda}_A+E_\mu^m
             -i(\lambda_A\sigma^m\bar{\psi}_{A\mu}
                 -\psi_{A\mu}\sigma^m\bar{\lambda}_A) \\
         &\quad +\frac{1}{4}\gamma_\mu^{rs}
              \lambda_A(\sigma^m\bar{\sigma}_{rs}+\sigma_{rs}\bar{\sigma}^m)
              \bar{\lambda}_A.
\end{split}
\end{equation}
As is presupposed in Eq.\eqref{NGtr2}, the coordinate transformation
induced by the $G$ transformation is given by $x'{}^\mu = x^\mu+\Delta x^\mu$.
This leads to the following transformation law of the space-time coordinate differential:
\begin{eqnarray}
  dx'{}^\mu=dx^\nu G_\nu^\mu(x), \quad G_\nu^\mu(x)=\frac{\partial x'{}^\mu}{\partial x^\nu}.
 \end{eqnarray}
This expression, together with the transformation Eq.\eqref{Maurertr},
yields the transformation law of the "vierbein" as
\begin{eqnarray}
  e_\mu'{}^m(x') &=& G_\mu^{-1\nu}(x)e_\nu^n(x)\Lambda_n^m(\alpha(x)).
 \end{eqnarray}
We can also define the "metric tensor" of extended local SUSY by means
of the "vierbein" as $g_{\mu\nu} = e_\mu^m \eta_{mn} e_\nu^n$, whose
transformation law is obtained as
\begin{eqnarray}
 g'_{\mu\nu}(x') = G_\mu^{-1\rho}(x)g_{\rho\sigma}(x)G_\nu^{-1\sigma}(x).
 \end{eqnarray}

The above preparation reveals that the construction of invariant actions
under the local $G$ transformation can be performed, with the aid of locally
covariant Maurer-Cartan 1-form, in the same way as the construction 
of invariant actions under the general coordinate transformation. 
Namely, we can write such invariant actions as
\begin{eqnarray}
  \Gamma=\int d^4x\mbox{ det }e(x)\mathcal{L}(x),
\end{eqnarray}
with $\mathcal{L}'(x')=\mathcal{L}(x)$, since $d^4x'=d^4x\mbox{ det}G$
and thus $d^4x'\mbox{ det }e'(x')=d^4x\mbox{ det }e(x)$.

We can construct invariant actions by using covariant derivatives
of locally covariant Maurer-Cartan 1-form. The covariant derivatives of
the components,
$\omega_{QA\alpha} = dx^\nu\omega_{QA\nu\alpha}$ and
$\bar{\omega}_{\bar{Q}A}^{\dot{\alpha}}=dx^\nu\bar{\omega}_{\bar{Q}A\nu}^{\dot{\alpha}}$,
are given by
\begin{equation}
\begin{split}
  \nabla_\mu\omega_{QA\nu\alpha} 
   &= \partial_\mu\omega_{QA\nu\alpha}
       +\frac{i}{4}\gamma_\mu^{mn}(\sigma_{mn})_\alpha^{\,\,\,\,\beta}\omega_{QA\nu\beta}
       -\Gamma^\rho_{\mu\nu}\omega_{QA\rho\alpha}, \\
  \nabla_\mu\bar{\omega}_{\bar{Q}A\nu}^{\dot{\alpha}} 
   &= \partial_\mu\bar{\omega}_{\bar{Q}A\nu}^{\dot{\alpha}}
       +\frac{i}{4}\gamma_\mu^{mn}(\bar{\sigma}_{mn})^{\dot{\alpha}}_{\,\,\,\,\dot{\beta}}
        \bar{\omega}_{\bar{Q}A\nu}^{\dot{\beta}}
       -\Gamma^\rho_{\mu\nu}\bar{\omega}_{\bar{Q}A\rho}^{\dot{\alpha}},
\end{split}
\end{equation}
where
\begin{eqnarray}
  \Gamma_{\sigma\rho}^\nu=e_n^{\nu}\partial_\rho e_\sigma^n
                      -e_n^{\nu}\gamma_{\rho}^{nr}e_\sigma^s\eta_{rs}
 \end{eqnarray}
is the "affine connection" in the extended supergravity.

Moreover, we are led to define the "Riemann curvature tensor" in the
same way as that in the case of general relativity:
\begin{eqnarray}
  R_{\sigma\mu\nu}^\rho=\partial_\nu\Gamma_{\sigma\mu}^\rho
                   -\partial_\mu\Gamma_{\sigma\nu}^\rho
                   +\Gamma_{\sigma\mu}^\lambda\Gamma_{\lambda\nu}^\rho
                   -\Gamma_{\sigma\nu}^\lambda\Gamma_{\lambda\mu}^\rho.
\end{eqnarray}
We can contract indices of the "Riemann curvature tensor"
to obtain the "Ricci tensor," $ R_{\mu\nu} = R_{\mu\nu\rho}^\rho$,
and the "scalar curvature," $R = g^{\mu\nu}R_{\mu\nu}$.

With the aid of all these ingredients, we finally arrive at invariant
actions of nonlinearly realized extended supergravity.
In particular, the minimal action without the fields for central charges
that has only quadratic terms of the components in the locally covariant
Maurer-Cartan 1-form is given by
\begin{equation}
\begin{split}
  \Gamma 
  &= \int d^4x\mbox{ det }e
      \bigg\{\Lambda-\frac{1}{2}R
                    +\epsilon^{\mu\nu\rho\sigma}
                     \omega_{QA\mu}\sigma_\sigma\nabla_\rho
                     \bar{\omega}_{\bar{Q}A\nu} \\
  &\quad \phantom{\int d^4x\mbox{ det }e\bigg\{}         
      +\frac{i}{2}m_{\frac{3}{2}AB}
       [\omega_{QA\mu}^\alpha\sigma_\alpha^{\mu\nu\beta}\omega_{QB\nu\beta}
        +\bar{\omega}_{\bar{Q}A\mu\dot{\alpha}}\bar{\sigma}^{\mu\nu\dot{\alpha}}_{\quad\,\,\dot{\beta}}
         \bar{\omega}_{\bar{Q}B\nu}^{\dot{\beta}}] \bigg\},
 \end{split}
\end{equation}
where $\Lambda$ is the cosmological constant. Note that the
normalizations of kinetic terms can be so chosen without loss of
generality by field rescalings.

When we adopt the unitary gauge, $\lambda_{A\alpha}=\bar{\lambda}_{A\dot{\alpha}}=0$,
the components of locally covariant Maurer-Cartan 1-form become
\begin{equation}
\begin{split}
  & \omega^m=dx^m+E^m=dx^\mu e^m_\mu, \\
  & \omega_{QA}^\alpha=\psi_A^\alpha, \quad
    \bar{\omega}_{\bar{Q}A\dot{\alpha}}=\bar{\psi}_{A\dot{\alpha}}, \quad
    \omega_M^{mn}=\gamma^{mn},
 \end{split}
\end{equation}
and the minimal action reduces to
\begin{equation}
\begin{split}
  \Gamma 
  &= \int d^4x\mbox{ det }e
      \bigg\{\Lambda-\frac{1}{2}R
                    +\epsilon^{\mu\nu\rho\sigma}\psi_{A\mu}
                     \sigma_\sigma \nabla_\rho\bar{\psi}_{A\nu} \\
  &\quad \phantom{\int d^4x\mbox{ det }e\bigg\{}         
      +\frac{i}{2}m_{\frac{3}{2}AB}[\psi_{A\mu}\sigma^{\mu\nu}\psi_{B\nu}
        +\bar{\psi}_{A\mu}\bar{\sigma}^{\mu\nu}\bar{\psi}_{B\nu}]\bigg\},
\end{split}
\end{equation}
where $m_{\frac{3}{2}AB}$ are gravitino masses.
As has been anticipated, we now see that $E^m$
is none other than the graviton,
$\psi_A^\alpha$ are the gravitinos, and $\gamma^{mn}$ is the spin connection.

We can also consider minimal actions with the fields for central
charges. For example, an action for the $\mathcal{N} = 2$ case is given by
\begin{equation}
\begin{split}
  \Gamma &= \int d^4x\mbox{ det }e\bigg[\Lambda-\frac{1}{2}R
             +\epsilon^{\mu\nu\rho\sigma}\omega_{QA\mu}\sigma_\sigma
               \nabla_\rho\bar{\omega}_{\bar{Q}A\nu}  \\
         &\quad \phantom{\int d^4x\mbox{ det }e\bigg[}
             +\frac{i}{2}m_{\frac{3}{2}AB}\{\omega_{QA\mu}\sigma^{\mu\nu}\omega_{QB\nu}
             +\bar{\omega}_{\bar{Q}A\mu}\bar{\sigma}^{\mu\nu}\bar{\omega}_{\bar{Q}B\nu}\} \\
         &\quad \phantom{\int d^4x\mbox{ det }e\bigg[}
             -\frac{M_X^2}{4}\{\nabla_\mu(\omega_{X\nu}
             +\bar{\omega}_{\bar{X}\nu})-\nabla_\nu(\omega_{X\mu}+\bar{\omega}_{\bar{X}\mu})\}^2 \\
              &\quad \phantom{\int d^4x\mbox{ det }e\bigg[}
             +\frac{M_X^2}{2}m_X^2(\omega_{X\mu}+\bar{\omega}_{\bar{X}\mu})^2 \\
         &\quad \phantom{\int d^4x\mbox{ det }e\bigg[}
             +f_{AB}M_X\epsilon^{\mu\nu\rho\sigma}\omega_{QA\mu}
              \sigma_\sigma(\omega_{X\rho}
             +\bar{\omega}_{\bar{X}\rho})\bar{\omega}_{\bar{Q}B\nu}\bigg],
\end{split}
\end{equation}
where the components associated with the central charges
of locally covariant Maurer-Cartan 1-form are denoted by
 \begin{equation}
 \begin{split}
  \omega_{X\mu}
   &= \frac{i}{2M_{s1}^2M_{s2}^2}\lambda_1\partial_\mu\lambda_2
       +\frac{1}{M_X^2}\partial_\mu C
       +\frac{i}{2\sqrt{2}}\bigg(\frac{1}{M_{s1}^2}\lambda_1\psi_{2\mu}
       -\frac{1}{M_{s2}^2}\lambda_2\psi_{1\mu}\bigg) \\
   &\quad +\frac{1}{M_X}A_\mu
       +\frac{1}{16M_{s1}^2M_{s2}^2}\gamma_{mn\mu}\lambda_1\sigma^{mn}\lambda_2, \\
  \bar{\omega}_{\bar{X}\mu}
   &= \frac{i}{2M_{s1}^2M_{s2}^2}\bar{\lambda}_1\partial_\mu\bar{\lambda}_2
       +\frac{1}{M_X^2}\partial_\mu\bar{C}
       +\frac{i}{2\sqrt{2}}\bigg(\frac{1}{M_{s1}^2}\bar{\lambda}_1
        \bar{\psi}_{2\mu}
       -\frac{1}{M_{s2}^2}\bar{\lambda}_2\bar{\psi}_{1\mu}\bigg) \\
   &\quad +\frac{1}{M_X}\bar{A}_\mu
       +\frac{1}{16M_{s1}^2M_{s2}^2}\gamma_{mn\mu}\bar{\lambda}_1\bar{\sigma}^{mn}
        \bar{\lambda}_2,
\end{split} 
\end{equation}
and the corresponding covariant derivative is given
by $\nabla_\mu\omega_{X\nu} = \partial_\mu\omega_{X\nu}
-\Gamma_{\mu\nu}^\rho\omega_{X\rho}$. The constants $M_X$ and $M_{sA}$
are introduced here for normalization.

When we adopt the unitary gauge, $\lambda_{A}=\bar{\lambda}_{A}=C=\bar{C}=0$, the action reduces to
\begin{equation}
\begin{split}
  \Gamma &= \int d^4x\mbox{ det }e\bigg[\Lambda-\frac{1}{2}R
             +\epsilon^{\mu\nu\rho\sigma}\psi_{A\mu}\sigma_\sigma\nabla_\rho
              \bar{\psi}_{A\nu} \\
         &\quad \phantom{\int d^4x\mbox{ det }e\bigg[}
             +\frac{i}{2}m_{\frac{3}{2}AB}\{\psi_{A\mu}\sigma^{\mu\nu}\psi_{QB\nu} +\bar{\psi}_{A\mu}\bar{\sigma}^{\mu\nu}\bar{\psi}_{B\nu}\} \\
         &\quad \phantom{\int d^4x\mbox{ det }e\bigg[}
            -\frac{1}{4}(\nabla_\mu B_\nu-\nabla_\nu B_\mu)^2 +\frac{m_X^2}{2}B_\mu^2 \\
         &\quad \phantom{\int d^4x\mbox{ det }e\bigg[}
             +f_{AB}\epsilon^{\mu\nu\rho\sigma}\psi_{A\mu}
              \sigma_\sigma B_\rho\bar{\psi}_{B\nu}\bigg],
\end{split} 
\end{equation}
where $B_\mu\equiv A_\mu+\bar{A}_\mu$.
This action is equivalent to the one given in Ref.\cite{Ferrara:1983fi},
where a nonlinearly realized local $\mathcal{N} = 2$ action was
considered by
reconciling the nonlinear realization of $\mathcal{N} = 2$ global SUSY
with the $\mathcal{N} = 2$ supergravity multiplet so that the full
theory has $\mathcal{N} = 2$ local SUSY up to the second order in the
fields. Note that the $\mathcal{N} = 2$ supergravity multiplet has one
spin-1 gauge boson, while we can introduce two spin-1 gauge fields
associated with two central charges in general,
although we have omitted the $(\omega_{X\mu}-\bar{\omega}_{\bar{X}\mu})$
dependence in the above example.

\subsection{The interactions with matters}

Let us introduce interactions of gravitinos with matters.
We define the $G$ transformation of a matter field $M(x)$ by
\begin{eqnarray}
M'(x') = \tilde{h} M(x), \quad \tilde{h} = e^{\frac{1}{2} {\alpha}_{mn}(x)\tilde{M}^{mn}},
 \end{eqnarray}
where $\tilde{M}^{mn}=0$ for a scalar field,
\begin{eqnarray}
(\tilde{M}^{mn})^{\,\,\,\,\beta}_{\alpha} = \frac{1}{2}(\sigma^{mn})^{\,\,\,\,\beta}_{\alpha}, \quad (\tilde{M}^{mn})^{\dot{\beta}}_{\,\,\,\,\dot{\alpha}} = \frac{1}{2}(\sigma^{mn})^{\dot{\beta}}_{\,\,\,\,\dot{\alpha}},
\end{eqnarray}
for a fermion field $\psi_{\alpha}(x)$, $\bar{\psi}_{\dot{\alpha}}(x)$,
and so forth.
Then the covariant derivative of the matter field is given by
\begin{eqnarray}
\nabla_\mu M = \left( \partial_\mu + \frac{i}{2} \gamma_\mu^{mn} \tilde{M}^{mn} \right) M,
 \end{eqnarray}
and its transformation law reads
\begin{eqnarray}
(\nabla_\mu M)'(x') = \tilde{h} G^{-1\nu}_\mu\nabla_\nu M(x).
 \end{eqnarray}

In terms of such matter fields, we obtain invariant actions
under the local symmetry. Namely, the matter action is given by
\begin{eqnarray}
  \Gamma_{matter}=\int d^4x\mbox{ det }e \ \mathcal{L}_{matter},
 \label{matterL}
\end{eqnarray}
where
\begin{eqnarray}
\mathcal{L}_{matter} = \mathcal{L}_{matter}(M, \nabla_\mu M, \omega_{QA}, \bar{\omega}_{\bar{Q}A}, \nabla_\mu\omega_{QA}, \nabla_\mu\bar{\omega}_{\bar{Q}A}, e^m_\mu, R_{\mu\nu\rho\sigma}),
 \end{eqnarray}
with
$\mathcal{L}'_{matter}(x') = \mathcal{L}_{matter}(x)$.

When we may identify the above matter fields as the Standard Model
particles, we obtain a theory which contains gravitino degrees of
freedom as many as the
number of SUSY interacting with the Standard Model fields.
This serves as a starting point to experimental predictions
of extended SUSY structure, which may be realized in Nature.

\section{Goldstino Embedding}

In the previous section, we have constructed our actions of nonlinearly
realized extended supergravity,
which enables us to obtain experimental predictions of these theories. 

As a first step, let us restrict ourselves to high-energy processes where
we can ignore the gravitino masses. In these situations, we naively
expect that nonlinear realization of extended global SUSY might provide
a good approximation to the original theory with local SUSY.

In the case of simple SUSY, this guess goes through as expected \cite{Deser:1977uq,Casalbuoni}.
However, as we will see below, in the case of
extended SUSY, such simplification generically does not give
a good approximation to the full supergravity theory even for high-energy processes.
Consequently, we have to extract experimental predictions from
our actions for extended local SUSY, instead of the global theories
from the beginning.

\subsection{The nonlinear realization of extended global SUSY}

We first review nonlinear realization of extended global SUSY
based on recent works by
Nishino \cite{Nishino:2000tv} and Clark and Love \cite{Clark:2000rv}.
The goldstino fields are denoted
by $\lambda_{A\alpha} (x)$, $\bar{\lambda}_{A\dot{\alpha}} (x)$,
and their extended SUSY transformation laws read
\begin{eqnarray}
  \lambda'_{A\alpha}(x') 
   &=&  \lambda_{A\alpha}(x)+\sqrt{2}M_{sA}^2\xi_{A\alpha}, \\
  \bar{\lambda}'_{A\dot{\alpha}}(x') 
   &=& \bar{\lambda}_{A\dot{\alpha}}(x)+\sqrt{2}M_{sA}^2\bar{\xi}_{A\dot{\alpha}},
\end{eqnarray}
with
\begin{eqnarray}
  x'{}^\mu 
   = x^\mu+\frac{i}{\sqrt{2}M_{sA}^2}[\xi_A\sigma^\mu\bar{\lambda}_A(x)
                                -\lambda_{A}(x)\sigma^\mu\bar{\xi}_A],
\end{eqnarray}
where $M_{sA}$ are superficial SUSY breaking scales. The invariant action under the
above transformations is a generalization of the Akulov-Volkov effective
action, which is given by
\begin{eqnarray}
  \Gamma_{AV}= \int d^4x \,\mathcal{L}_{AV}, \quad \mathcal{L}_{AV} =  - M_s^4 \mbox{ det } \left(\delta^\nu_\mu - \frac{i}{M_{sA}^4} \lambda_A \overleftrightarrow{\partial_\mu} \sigma^\nu\bar{\lambda}_A \right).
 \end{eqnarray}

Although we can introduce such superficial SUSY-breaking scales as many as the number
of SUSY in the action, the goldstino field redefinitions, 
\begin{eqnarray}
 \lambda_A \to \frac{M_{sA}^2}{M_s^2}\lambda_A,
 \end{eqnarray}
render the above Lagrangian into
\begin{eqnarray}
 \mathcal{L}_{AV} =  - M_s^4 \mbox{ det } \left(\delta^\nu_\mu - \frac{i}{M_{s}^4} \lambda_A \overleftrightarrow{\partial_\mu} \sigma^\nu\bar{\lambda}_A \right),
 \end{eqnarray}
where $M_{s}$ is the common SUSY-breaking scale.
This shows that physically
only one SUSY-breaking scale is present in the action of nonlinear
realization with extended global SUSY.

\subsection{Comparison with the Deser-Zumino construction}

We here turn to identify goldstino degrees of freedom in the
Deser-Zumino construction \cite{Deser:1977uq} with those in
our construction and compare ours
to the nonlinear realization for the global case.
 
The minimal action with nonlinearly realized extended supergravity 
is approximately given by
\begin{equation}
\begin{split}
  \Gamma &\simeq \int d^4x\mbox{ det }e\bigg[\L
                  -\frac{1}{2}R
                  +\epsilon^{\mu\nu\rho\sigma}\omega_{QA\mu}
                   \sigma_\sigma\partial_\rho\bar{\omega}_{\bar{Q}A\nu} \\
         &\quad \phantom{\int d^4x\mbox{ det }e\bigg[}
                  +\frac{i}{2}m_{\frac{3}{2}AB}
                   \{\omega_{QA\mu}^\alpha\sigma_\alpha^{\mu\nu\beta}\omega_{QB\nu\beta}
                     +\bar{\omega}_{\bar{Q}A\mu\dot{\alpha}}
                      \bar{\sigma}^{\mu\nu\dot{\alpha}}_{\quad\,\dot{\beta}}
                      \bar{\omega}_{\bar{Q}B\nu}^{\dot{\beta}}\}\bigg],
 \end{split}
\end{equation}
where
\begin{equation}
 \begin{split}
  \omega_{QA\mu}^\alpha 
   &= \frac{\sqrt{2}}{M_{sA}^2}\partial_\mu\lambda_A^\alpha
       +\psi_{A\mu}^\alpha, \\
  \bar{\omega}_{\bar{Q}A\mu\dot{\alpha}} 
   &= \frac{\sqrt{2}}{M_{sA}^2}\partial_\mu\bar{\lambda}_{A\dot{\alpha}}
       +\bar{\psi}_{A\mu\dot{\alpha}}, \\
  e_\mu^m &= \delta_\mu^m+E_\mu^m-\frac{i}{M_{sA}^4}\lambda_A\sigma^m
             \overleftrightarrow{\partial}_\mu\bar{\lambda}_A
             -\frac{i}{\sqrt{2}M_{sA}^2}
              (\lambda_A\sigma^m\bar{\psi}_{A\mu}
               -\psi_{A\mu}\sigma^m\bar{\lambda}_A). \\
 \end{split}
\end{equation}
Here, we have reduced the action up to the second order in the fields
and introduced normalization factors $M_{sA}$.
We further set the cosmological constant and the non-diagonal components of
gravitino masses%
\footnote{They imply a possibility of gravitino oscillation.}
to be vanishing, that is,
$\Lambda=0$ and $m_{\frac{3}{2}AB}=0$ for $A \neq B$, which results in
\begin{equation}
\begin{split}
  \Gamma 
   &\simeq \int d^4x\bigg[
            -\frac{1}{2}R  +\epsilon^{\mu\nu\rho\sigma}\psi_{A\mu}\sigma_\sigma\partial_\rho
             \bar{\psi}_{A\nu} \\
   &\quad \phantom{\int d^4x\bigg[}   
            +\frac{i}{2}m_{\frac{3}{2}A}\psi_{A\mu}^\alpha\sigma_\alpha^{\mu\nu\beta}\psi_{A\nu\beta}
+\frac{i}{2}m_{\frac{3}{2}A}\bar{\psi}_{A\mu\dot{\alpha}}
             \bar{\sigma}^{\mu\nu\dot{\alpha}}_{\quad\,\,\dot{\beta}}
             \bar{\psi}_{A\nu}^{\dot{\beta}} \\
   &\quad \phantom{\int d^4x\bigg[}
            +\frac{\sqrt{2}m_{\frac{3}{2}A}}{M_{sA}^2}i\partial_\mu
             \lambda_A^\alpha\sigma_\alpha^{\mu\nu\beta}\psi_{A\nu\beta}+\frac{\sqrt{2}m_{\frac{3}{2}A}}{M_{sA}^2}i\partial_\mu
             \bar{\lambda}_{A\dot{\alpha}}\bar{\sigma}^{\mu\nu\dot{\alpha}}_{\quad\,\,\dot{\beta}}
             \bar{\psi}_{A\nu}^{\dot{\beta}}\bigg].
 \end{split}
\end{equation}
This form has the local SUSY within the present approximation.

The above expression seems to have no goldstino kinetic terms, though it
has mixed kinetic terms of goldstinos and gravitinos. Thus, we are led
to redefine the gravitino fields so that the action has the canonical
goldstino kinetic terms without the mixed terms. The redefinition
of the gravitino fields is given by
\begin{equation}
\begin{split}
  \psi_A^\mu &\rightarrow \psi_A^\mu+iC_A\sigma^\mu\bar{\lambda}_A, \\
  \bar{\psi}_A^\mu &\rightarrow \bar{\psi}_A^\mu-iC^{\ast}_A\bar{\sigma}^\mu\lambda_A,
\end{split}
 \end{equation}
where $C_A$ are constants to be determined below.
This makes the action to be
\begin{equation}
\begin{split}
  \Gamma 
   &\simeq \int d^4x\bigg[
            -\frac{1}{2}R +\epsilon^{\mu\nu\rho\sigma}\psi_{A\mu}\sigma_\sigma\partial_\rho
             \bar{\psi}_{A\nu} \\
            &\quad \phantom{\int d^4x\bigg[}  +\frac{i}{2}m_{\frac{3}{2}A}\psi_{A\mu}\sigma^{\mu\nu}\psi_{A\nu}
+\frac{i}{2}m_{\frac{3}{2}A}\bar{\psi}_{A\mu}
             \bar{\sigma}^{\mu\nu}
             \bar{\psi}_{A\nu} \\
   &\quad \phantom{\int d^4x\bigg[}  +\bigg(
            i\frac{3\sqrt{2}C_A m_{\frac{3}{2}A}}{M_{sA}^2}
            -i\frac{3\sqrt{2}C^{\ast}_A m_{\frac{3}{2}A}}{M_{sA}^2}+6|C_A|^2\bigg)i\lambda_A\sigma^\mu\partial_\mu\bar{\lambda}_A \\
   &\quad \phantom{\int d^4x\bigg[}
            +\left(2C_A+\frac{\sqrt{2}m_{\frac{3}{2}A}}{M_{sA}^2} \right)i\partial_\mu
             \lambda_A\sigma^{\mu\nu}\psi_{A\nu}
            +\left(2C_A^\ast+\frac{\sqrt{2}m_{\frac{3}{2}A}}{M_{sA}^2}\right)i\partial_\mu
             \bar{\lambda}_{A}\bar{\sigma}^{\mu\nu}\bar{\psi}_{A\nu} \\
&\quad \phantom{\int d^4x\bigg[}-\frac{3}{2}iC_Am_{\frac{3}{2}A}\psi_{A\mu}\sigma^{\mu}\bar{\lambda}_A+\frac{3}{2}iC^\ast_Am_{\frac{3}{2}A}\bar{\psi}_{A\mu}
             \bar{\sigma}^{\mu}
             \lambda_{A} -6|C_A|^2m_{\frac{3}{2}A} \left(\lambda_{A}\lambda_A + \bar{\lambda}_{A}\bar{\lambda}_A\right) \biggr].
\end{split} 
\end{equation}

We now choose $C_A$ so that the mixed kinetic terms of goldstinos and
gravitinos disappear, which is achieved by
\begin{eqnarray}
  C_A=-\frac{m_{\frac{3}{2}A}}{\sqrt{2}M_{sA}^2}. \label{ca}
 \end{eqnarray}
We also impose the condition that the action has the canonical goldstino
kinetic terms, which requires
\begin{eqnarray}
            6C_A^2 =1.
 \end{eqnarray}
Together with Eq.\eqref{ca}, this determines the normalization
factors $M_{sA}$ as physical scales:
\begin{eqnarray}
  m_{\frac{3}{2}A}=\frac{M_{sA}^2}{\sqrt{3}},
 \end{eqnarray}
which is a generalization of the usual relation Eq.(\ref{gravitinomass}).
Hence we have finally obtained a desired form
\begin{equation}
\begin{split}
  \Gamma 
   \simeq \int d^4x\bigg[
            &-\frac{1}{2}R +
  \epsilon^{\mu\nu\rho\sigma}\psi_{A\mu}\sigma_\sigma\partial_\rho
             \bar{\psi}_{A\nu}+i\lambda_A\sigma^\mu\partial_\mu\bar{\lambda}_A \\
             &+\frac{i}{2}m_{\frac{3}{2}A}\psi_{A\mu}\sigma^{\mu\nu}\psi_{A\nu}
+\frac{i}{2}m_{\frac{3}{2}A}\bar{\psi}_{A\mu}
             \bar{\sigma}^{\mu\nu}
             \bar{\psi}_{A\nu} \\
             &+i\frac{\sqrt{6}}{4}m_{\frac{3}{2}A}\left(\psi_{A\mu}\sigma^{\mu}\bar{\lambda}_A-\bar{\psi}_{A\mu} 
             \bar{\sigma}^{\mu} 
             \lambda_{A} \right)-m_{\frac{3}{2}A}\left(\lambda_{A}\lambda_A + \bar{\lambda}_{A}\bar{\lambda}_A\right) \biggr],
\end{split}\label{approximate}
\end{equation}
which coincides with the Deser-Zumino construction for ${\cal N}=1$.
Note that the matter interaction Eq.(\ref{matterL})
contains the physical SUSY-breaking scales $M_{sA}$
in the factor ${\rm det}\,e$, which determine
the universal interactions of the goldstinos with matter fields.

In the case of $\mathcal{N}=1$ SUSY,
the Akulov-Volkov global action may be regarded as the first approximation to
the full supergravity action since it is the starting point of the
Deser-Zumino construction.
However, when we consider the case of extended SUSY,
the story is different from the $\mathcal{N}=1$ case.
As we have seen in the previous subsection, one common SUSY-breaking scale
essentially appears in the action for nonlinear realization of
extended global SUSY.
In contrast, the above goldstino embedding in supergravity generically yields
physically independent SUSY-breaking scales as many as the number of SUSY.
Hence we conclude that 
the nonlinear realization of extended global SUSY 
does not properly approximate 
the nonlinearly realized extended supergravity 
theory even for high-energy processes.%
\footnote{One obvious exception is the case that there is only one
SUSY-breaking scale so that all the gravitino masses take
the same value in the original action from the start.}

\section{Conclusion}

In this paper, we have provided nonlinear realization
of extended supergravity by using coset construction.
One of the motivations to construct such a theory
is to serve for investigating the number of SUSY
by counting the number of gravitinos at the scale within experimental reach.

We also consider goldstino embedding in the locally invariant theory.
In the case of $\mathcal{N}=1$ SUSY, we may use the nonlinear
realization of global SUSY to extract approximate experimental
predictions of the local theory.
However, in the case of extended SUSY, we cannot use the nonlinear
realization of extended global SUSY, or rather,
we generically have to use the nonlinearly realized extended
supergravity action even for high-energy processes.

It seems intriguing to investigate characteristic experimental
signatures of theories with more than one kind of gravitinos.
Along the way, inclusion of linear SUSY multiplets into the present
formalism may be useful for further research in connection with
phenomenology under the setup of SUSY Standard Model.
Mirror superpartners other than multiple gravitinos such as mirror gauginos,
higgsinos, quarks and leptons might play crucial roles
in future particle phenomenology. As such, low-energy extended SUSY
structure is a candidate window into basic laws in Nature.

\section*{Acknowledgements}

This work was supported by the Grant-in-Aid for Yukawa International
Program for Quark-Hadron Sciences, the Grant-in-Aid
for the Global COE Program "The Next Generation of Physics,
Spun from Universality and Emergence", and
World Premier International Research Center Initiative
(WPI Initiative), MEXT, Japan.

%\section*{Appendix}

%\newpage

\end{document}